\documentclass[showpacs,showkeys,amssymb,aps,epsfig,graphicx,graphics,epsf]{revtex4}
\newcommand{\be}{\begin{equation}} \newcommand{\ee}{\end{equation}}
\newcommand{\bea}{\begin{eqnarray}}\newcommand{\eea}{\end{eqnarray}}

\usepackage{graphicx}

\begin{document}

\hfill{SINP/TNP/2009/24}
\vspace*{10mm}
\title{Turbulent flow  in graphene}
\author{Kumar S. Gupta\footnote{Email : kumars.gupta@saha.ac.in~ }}
\affiliation{ Theory Division, Saha Institute of Nuclear Physics, 1/AF Bidhannagar, Calcutta - 700064, India }

\author{Siddhartha Sen \footnote{Email : sen@maths.ucd.ie, Emeritus Fellow, TCD}}

\affiliation {School of Mathematical Sciences, UCD, Belfield, Dublin 4, Ireland}
\affiliation {Department of Theoretical Physics, Indian Association for the Cultivation of Science, Calcutta - 700032, India }

\begin{abstract}

We demonstrate the possibility of  turbulent flow of electrons in graphene in the hydrodynamic region, by calculating the corresponding turbulent probability density function. This is used to calculate the contribution of the turbulent flow to the conductivity within a quantum Boltzmann approach. The dependence of the conductivity on the system parameters arising from the turbulent flow is very different from that due to scattering. 

\end{abstract}

\pacs{47.27.-i, 05.60.Gg, 73.23.-b, 81.05.Uw }
\keywords{Turbulent flow, Graphene}
\maketitle

\vspace*{0.5cm}


Graphene continues to offer unexpected possibilities for theoretical studies which are novel and can be experimentally tested. In recent work \cite{muller1}, a regime has been identified for the electric transport in an ideal single layer undoped graphene sheet at zero applied gate voltage, where a hydrodynamic description using a quantum Boltzmann equation is relevant. The ratio of the shear viscosity $\eta$ to the entropy density $s$ in this region was shown to be very small \cite{muller2}. This makes graphene behave like a nearly perfect fluid. It was also suggested that turbulent electronic flow in this regime might be possible, since a hydrodynamic system with low viscosity is expected to have a quantum turbulent flow \cite{book}. A turbulent regime is characterized by a large value for the Reynold's number, which for graphene at a given temperature $T$ is given by \cite{muller2}
$$
Re = \frac{s/k_B}{\eta / \hbar} \frac{k_B T}{ {\hbar v} / L} \frac{u_{typ}}{v},
$$
where $v$ is the Fermi velocity, $u_{typ}$ denotes a typical velocity and $L$ is the characteristic length scale for the velocity gradients. Furthermore it was suggested in \cite{muller2} that for $Re \sim 10^2$, turbulent flow might be possible in graphene.
Thus determining the turbulent probability density function for graphene and using it to calculate its contribution to the conductance in graphene is of interest. These two issues are addressed in their Letter.

Quantum turbulence is an approximate way to understand scaling behaviour in a quantum fluid in a nonequilibrium state. The framework for understanding the similarities and differences between classical fluid turbulence and quantum fluid turbulence is based on a nonlinear Schrodinger equation model \cite{gross}  or a filament model \cite{sch}. The theoretical approach we use may be viewed as a generalization of the nonlinear Schrodinger equation approach. It has been stressed by Zakharov et al \cite{book} that turbulence is a physical phenomenon which is much wider in scope than simply fluid turbulence. A quantum turbulent state is possible for a complex quantum system with many degrees of freedom. It represents a highly excited state of the quantum system where energy is pumped in one scale and dissipated due to nonlinearities present at a different scale.  Graphene is a complex system described by a nonlinear quantum Hamiltonian given by Eqn. (1). Furthermore, recent work h
 as established that this system has a hydrodynamic regime with a high Reynold's number. In view of this we would like to follow the well established methods \cite{book,n1,n2} to determine the turbulent PDF for graphene and calculate the contribution made by this component to conductance.

Turbulent flows are characterized by a probability distribution function (PDF) with scaling properties and an associated nonvanishing energy flux. Quantum turbulent flows have been studied using the methods of quantum field theory, as such systems have many degrees of freedom exhibiting stochastic behaviour \cite{mark}. Systems where the wave motion is a recognized feature and turbulent flows are generated due to nonlinear interactions are known as weak wave turbulent systems \cite{book,mark}.  

In this Letter we calculate a possible turbulent flow probability distribution function (TPDF) for graphene in the hydrodynamic regime treating it as a weak wave system. The wave motion comes from the free Dirac equation. The nonlinearity in graphene is due to the electron-electron Coulomb interaction, which is described by a term quartic in the field operators \cite{muller1}.  We show that these combined types of interactions admit a solution with a turbulent flow. Using the TPDF we then calculate the conductance due to this flow. We find that the contribution to the conductance due to turbulent flow has a structure different from that found using scattering \cite{muller1}. 


In Ref. \cite{muller1} it was shown that graphene near the Fermi points is described by a Hamiltonian given by
\bea \label{1}
H &=& H_0 + H_1 , \nonumber\\
H_0 &=& \sum_{\lambda,i} \int \frac{d^2 k}{(2 \pi)^2} \lambda v_F k a^{\dagger}_{\lambda i}({\bf k}) a_{\lambda i}({\bf k}) , \nonumber \\
H_1 &=& \sum_{\lambda_1,\lambda_2,\lambda_3,\lambda_4; i,j} 
\int \frac{d^2 k_1}{(2 \pi)^2} \frac{d^2 k_2}{(2 \pi)^2} \frac{d^2 k_3}{(2 \pi)^2} 
\frac{d^2 k_4}{(2 \pi)^2} \delta^2({\bf k_1} + {\bf k_2} - {\bf k_3} - {\bf k_4}) \nonumber \\
&& \times T_{\lambda_1 \lambda_2 \lambda_3 \lambda_4} ({\bf k_1}, {\bf k_2}, {\bf k_3}, {\bf k_4}) a^{\dagger}_{\lambda_4 j}({\bf k_4}) 
a^{\dagger}_{\lambda_3 i}({\bf k_3} ) a_{\lambda_2 i}({\bf k_2}) a_{\lambda_1 j}({\bf k_1}),
\eea
where $a_{+i}, a_{-i}$ are the Fourier mode operators, $i,j = 1,...,4$ for graphene corresponding to two valleys and two spins and $\lambda = (+,-)$. We also define $k = | {\bf k}|$, which is the modulus of the two momentum ${\bf k}$. The explicit expression for $T_{\lambda_1 \lambda_2 \lambda_3 \lambda_4} ({\bf k_1}, {\bf k_2}, {\bf k_3}, {\bf k_4} )$ is given in Ref. \cite{muller1}. For our discussion, we need to note that under scaling,
\be
T(\mu {\bf k_1}, \mu {\bf k_2}, \mu {\bf k_3}, \mu {\bf k_4}) = \frac{1}{\mu} T({\bf k_1}, {\bf k_2}, {\bf k_3} , {\bf k_4}),
\ee
where $\mu$ is a scalar.

A turbulent flow represents a steady state of the system in which, unlike a thermal equilibrium state, there exists a transport of a charge current. Such flows are expected to have scaling properties. To calculate the PDF for such a flow, we follow the method described in Ref. \cite{book}, modified to deal with quantum massless Dirac system. The basic idea is to calculate
\be
\frac{d}{dt} n_{k,\lambda} = \frac{d}{dt} <a^{\dagger}_{k,\lambda} a_{k,\lambda} >
\ee
between two ``in-states'' \cite{mark}, which are eigenstates of $H_0$. Next we look for scaling solutions which have the property that 
\be
\frac{d}{dt} n_{k,\lambda} = 0
\ee
and have a nonvanishing associated flux, to be discussed later. From Eq. (4) we find that 
\bea
0 &=& \sum_{\lambda_1,\lambda_2,\lambda_3,\lambda_4; i,j} 
\int \frac{d^2 k_1}{(2 \pi)^2} \frac{d^2 k_2}{(2 \pi)^2} \frac{d^2 k_3}{(2 \pi)^2} 
 \delta^2({\bf k_1} + {\bf k_2} - {\bf k_3} - {\bf k}) 
  |T_{\lambda_1 \lambda_2 \lambda_3 \lambda_4} ({\bf k_1}, {\bf k_2}, {\bf k_3}, {\bf k})|^2 n_{k_1} n_{k_2} n_{k_3} n_k  \nonumber \\
&&  \times \left [
-16 \left ( \frac{1}{n_k} + \frac{1}{n_{k_3}} -  \frac{1}{n_{k_1}} - \frac{1}{n_{k_2}} \right ) 
+ 8   \left ( \frac{1}{n_{k_3} n_{k}} +  \frac{1}{n_{k_1} n_{k}} -  \frac{1}{n_{k_2} n_{k_3}} 
- \frac{1}{n_{k_1} n_{k_2}} \right ) \right ].
\eea
We would like to obtain the stationary solutions in terms of the energy instead of the momenta. We shall further assume that the solutions are isotropic, so that both $n_{k, \lambda}$ and the energy $\epsilon$ depend only on the magnitude $k$ of the momentum vector ${\bf k}$.
Following the procedure of Zakharov \cite{zakha1} we change variables from $k$ to $\epsilon (k)$, where the latter denotes the energy and replace $T$ by 
\be
U (\epsilon_1 \epsilon_2 \epsilon_3 \epsilon ) = (k_1 k_2 k_3 k) \left | \frac{d \epsilon_1}{d k_1} \frac{d \epsilon_2}{d k_2} \frac{d \epsilon_3}{d k_3} \frac{d \epsilon}{d k} \right |^{-1} \int |T_{k_1 k_2 k_3 k_4}|^2 \delta^2(k_1 + k_2 - k_3 - k) d \Omega_1 d \Omega_2 d \Omega_3 ,
\ee
where we have used the relation $d^2k_i = k_i d k_i d \Omega_i$. From Eq. (6) it can be easily seen that the quantity $U$ is invariant under scaling.

From Eq. (5) and Eq. (6) and using the free particle density, we get 
\bea 
0 &=& \int_0^{\infty}  d \epsilon_1 d \epsilon_2 d \epsilon_3 U (\epsilon_1 \epsilon_2 \epsilon_3 \epsilon ) \delta ( \epsilon_1 + \epsilon_2 - \epsilon_3 - \epsilon )
n_{k_1} n_{k_2} n_{k_3} n_k \nonumber \\
&& \times \left [
-16 \left ( \frac{1}{n_k} + \frac{1}{n_{k_3}} -  \frac{1}{n_{k_1}} - \frac{1}{n_{k_2}} \right ) 
+ 8   \left ( \frac{1}{n_{k_3} n_{k}} +  \frac{1}{n_{k_1} n_{k}} -  \frac{1}{n_{k_2} n_{k_3}} 
- \frac{1}{n_{k_1} n_{k_2}} \right ) \right ].
\eea
We now analyze Eq. (7). The $\delta$-function can be used to carry out the $\epsilon_3$ integral. What remains is a region $D$ of the $(\epsilon_1,\epsilon_2)$ plane. In addition we have a constraint that $\epsilon_3 = \epsilon_1 + \epsilon_2 - \epsilon \geq 0$. Following  \cite{zakha1}, we divide D into four sectors as follows 
\bea
D_1 &=& \{(\epsilon_1, \epsilon_2) \in D ~|~ \epsilon_1 < \epsilon, ~ \epsilon_2 < \epsilon \} \nonumber \\
D_2 &=& \{(\epsilon_1, \epsilon_2) \in D ~|~ \epsilon_1 > \epsilon, ~ \epsilon_2 > \epsilon \} \nonumber \\
D_3 &=& \{(\epsilon_1, \epsilon_2) \in D ~|~ \epsilon_1 < \epsilon, ~ \epsilon_2 < \epsilon \} \nonumber \\
D_4 &=& \{(\epsilon_1, \epsilon_2) \in D ~|~ \epsilon_1 > \epsilon, ~ \epsilon_2 < \epsilon \} 
\eea
It was shown in \cite{zakha1} that the regions $D_2, D_3, D_4$ can be mapped into $D_1$ using the Zakharov transformations. For example $D_2$ with variables $(\epsilon^{\prime}_1, \epsilon^{\prime}_2)$ can be mapped to $D_1$ with variables $(\epsilon_1, \epsilon_2)$ using
\bea
\epsilon^{\prime}_1 &=& \frac{\epsilon \epsilon_1}{\epsilon_1 + \epsilon_2 - \epsilon} \nonumber \\
\epsilon^{\prime}_2 &=& \frac{\epsilon \epsilon_2}{\epsilon_1 + \epsilon_2 - \epsilon}
\eea and so on. Using these transformations and the ansatz $n(\epsilon) = C \epsilon^{-x}$, we get
\be
0 = \int_{D_1} d \epsilon_1 d \epsilon_2 U (\epsilon_1, \epsilon_2, \epsilon_1 + \epsilon_2 - \epsilon, \epsilon) (\epsilon_1 \epsilon_2 (\epsilon_1 + \epsilon_2 - \epsilon) \epsilon)^x 
\left [ 1 + (\frac{\epsilon_1 + \epsilon_2 - \epsilon}{\epsilon})^{y} - (\frac{\epsilon_2}{\epsilon})^{y} - (\frac{\epsilon_1}{\epsilon})^{y} \right ] \times {\rm {other ~ factors}},
\ee
where $y = 3x - 3$. The factor within the square brackets in Eq.(10) can be made to vanish by choosing $y=0$ or $y=1$, irrespective of other factors in the equation.  It is possible to show \cite{book,sanyal} that only the solution corresponding to $y=0$ or $x=1$ has an associated nonzero flux and hence represents a turbulent state. This is the solution that we shall use.

Our solution for the TPDF is thus given by
\be
n(\epsilon) = \frac{C}{\epsilon}= \frac{C}{\hbar k},
\ee
where for the moment we have set $v_F=1$. We now need to determine $C$. To do this, we observe that the total number of turbulent particles per unit area, that is the turbulent number density $N_T$ is given by
\be
N_T = \frac{C}{2 \pi} \frac{k_{max}}{\hbar},
\ee
We find
\be
C \epsilon_{max} = N_T v^2_F \left [ \frac{(2 \pi \hbar)^2}{4} \frac{1}{2 \pi} \right ]^{-1},
\ee
where $\epsilon_{max} = v_F k_{max} = \gamma k_B T$, where $T$ is the background temperature of the system before the turbulent flow starts, $k_B$ 
is the Boltzmann constant and $\gamma << 1$ is a constant, which is required for the hydrodynamic regime to be relevant.


We now proceed to calculate the conductance using the quantum Boltzmann equation. For this purpose we introduce a time dependent electric field ${\bf E}(t)$, which is assumed to be small. We write the time dependent density
\be
f_\lambda = <a^{\dagger}_{\lambda, i} (k, t) a_{\lambda, i} (k, t) >
\ee
and $f^{0}_\lambda = \frac{C}{k}$, the turbulent time independent PDF and determine $f_\lambda$ by using the equation
\be
\left ( \frac{\partial}{\partial t} + {\bf E}.\frac{\partial}{\partial {\bf k}} \right ) f_{\lambda} = 0.
\ee
We use the ansatz
\be
\tilde{f}_\lambda(k, \omega) = \tilde{f}^0_\lambda(k) + \frac{{\bf E}.{\bf k}}{k} 
\tilde{g}_\lambda(k, \omega),
\ee
where
\be
\int \tilde{f}_\lambda(k, \omega) e^{i \omega t} \frac{d \omega}{2 \pi} = f_{\lambda}(k, t).
\ee
We find 
\be
\tilde{f}_\lambda = f^0_{\lambda} + \frac{{\bf E}.{\bf k}}{k} 
\left ( \frac{\partial f^0_{\lambda}}{\partial k} \right ) \frac{1}{(-i \omega + 0_+)},
\ee
where we have ignored the terms higher order in ${\bf E}$. It is now a simple matter to calculate the conductance. Taking the electric field to be in the $x$ direction, we have 
\be
\sigma_T (\omega) = \frac{<J_1>}{E_x},
\ee
where $(<J_1>) = f_{\lambda}$ and $\sigma_T$ is the contribution of the turbulent flow to the conductivity. Substituting for $f_{\lambda}$ and introducing back the Fermi velocity $v_F$, we get
\be
 \sigma_T (\omega) = \frac{e^2}{h} \frac{N_T v_F^2}{2\beta k_B T} 
 \log \left ( \frac{\gamma k_B T L}{v_F \hbar} \right ) 
\left ( \frac{1}{-i \hbar\omega + 0^+} \right ) 
\ee 
where $v_F k = \gamma k_B T,~ \gamma <<1$ and L is the typical sample size. It may be noted that the spin valley degeneracy factor does not appear in the contribution to the conductivity coming from the turbulence flow. It is useful to compare the conductivity $\sigma_T$ obtained above with the expression for conductivity due to scattering $\sigma_{\rm sc}$ obtained in \cite{muller1}, which is given by
\be
\sigma_{\rm sc} = \frac{e^2}{h} \frac{N k_B T ln2}{-i \hbar \omega + \kappa k_B T \alpha^2},
\ee
where $\alpha$ is the electron-electron interaction strength and $N$ denotes the spin valley degeneracy in graphene. Comparing the expressions of $\sigma_T$ obtained here with that of $\sigma_{\rm sc}$ obtained in \cite{muller1} we see that they have very different dependence on the physical parameters of the system. In particular, $\sigma_T (\omega)$ is independent of the spin-valley degeneracy in graphene. The temperature dependence of $\sigma_T$ and $\sigma_{\rm sc}$ are also quite different.

To summarize, we have calculated the contribution of the turbulent flow to the conductance in graphene. The dependence of the conductivity on the system parameters is very different from that described in recent calculations \cite{muller1} using scattering. Note that the temperature in our formula is a natural cutoff describing the hydrodynamic region. It represents the background equilibrium state temperature of the graphene sample. The factor $N_T$ present in the expression of the conductance is assumed to be of the same order of magnitude as the uniform electron density in the system. This corresponds to the physical picture of a finite macroscopic fraction of the electron fluid exhibiting turbulent behaviour.

The turbulent flow discussed here is a novel consequence of nonlinear electron-electron interactions in graphene. The turbulent flow found has superfluid like properties in the sense that the dispersion relation is linear in the momentum and the system has a macroscopic turbulent flow density. The linear dispersion implies that dissipation through scattering with a defect is only possible when the fluid velocity is greater than $v_F$. The absence of dissipation implies that the electron flow is frictionless in the turbulent regime.

As emphasized in \cite{muller1}, in order to experimentally access this region, a large graphene sample is required. With a suitable sample, the contribution of the conductivity
 due to the turbulent flow might be identified through its particular dependence on the system parameters.


\end{document}